# Re-entrant spin-glass freezing and magneto-dielectric behaviour of Li$_3$NiRuO$_5$, a layered rock-salt related oxide

**Sanjay Kumar Upadhyay,[1] Kartik K Iyer,[1] S. Rayaprol,[2] V. Siruguri,[2] and E.V. Sampathkumaran[1]**

[1]Tata Institute of Fundamental Research, Homi Bhabha Road, Colaba, Mumbai 400005, India

[2]UGC-DAE Consortium for Scientific Research Mumbai Centre, R-5 Shed, BARC Campus, Trombay, Mumbai – 400085, India

We report the results of neutron diffraction, *ac* and *dc* magnetization, heat-capacity, complex dielectric permittivity and pyrocurrent measurements on an oxide, Li$_3$NiRuO$_5$, hitherto not paid much attention in the literature, except for a previous report on its promising electrochemical performance. We emphasize on the following findings: (i) Observation of re-entrant spin-glass behaviour; that is, this oxide undergoes ferrimagnetic ordering below 80 K, entering spin-glass regime around 12 K. (ii) There is no prominent feature in the complex dielectric permittivity (in particular, at the magnetic transitions) in the absence of external magnetic field, indicative of the absence of ferroelectricity. However, there is distinct evidence for magneto-dielectric (MDE) coupling. The sign of MDE coupling also changes as the sample is cooled from ferrimagnetic state to spin-glass regime. (iii) There are pyroelectric anomalies in the vicinity of 30 – 70 K, presumably arising from thermally stimulated depolarization current.

(This will appear in print in Journal of Materials Chemistry C)

## Introduction

In recent years, the synthesis of layered oxides, $Li_3M_2RuO_6$ [1] and $Li_3MRuO_5$ (M= Co, Ni) [2], as well as $Li_3MRuO_5$ (M= Mn, Fe) [3] based on $LiCoO_2$ and $Li_2MnO_3$ [4, 5] relevant to Li-ion batteries, have been reported. These oxides have been studied from the electrochemical angle by Laha et al [1-3]. Considering that detailed magnetic behaviour of these oxides had not been known, we have initiated a systematic investigation of magnetic properties on these oxides, particularly to explore the role of intrinsic crystallographic disorder in the transition metal layer. We consider it important to understand dielectric and pyroelectric current ($I_{pyro}$) behaviour of these oxides as well, as such studies may bear relevance to the currently active field of multiferroic and magnetodielectric (MDE) coupling. With these motivations, we have undertaken the oxide, $Li_3NiRuO_5$, for magnetic, dielectric and pyroelectric current studies as a function of temperature ($T$). This oxide has been reported to crystallize in a rhombohedral crystal structure (space group $R\overline{3}m$) by Laha et al [2] on the basis of x-ray diffraction results. Our preliminary neutron diffraction results show that this compound might be monoclinic in structure, possibly in the sub-group $P2/m$. Based on preliminary dc magnetic susceptibility ($\chi$) studies, the compound was proposed to order ferrimagnetically below 80 K. We report here that this oxide undergoes re-entrant spin-glass transition and exhibits MDE coupling and $I_{pyro}$ anomaly without ferroelectricity.

## Experimental details

### Synthesis

The polycrystalline $Li_3NiRuO_5$ was prepared through conventional solid state route [3]. Initial materials used for the synthesis were $Li_2CO_3$, $NiC_2O_4.2H_2O$ and $RuO_2$, with purity better than 99.9 %. Stoichiometric amounts of all constituents were mixed thoroughly, followed by the calcination at 810 $^0$C for 12 h. This heat-treated specimen was pelletized and sintered at $925^0$C for 24 h.

### Structural characterization

X-ray diffraction (XRD) measurement was done with Cu $K_\alpha$ radiation (Xpert Pro MPD, Panalytical) and the powder neutron diffraction measurements were carried out on the UGC-DAE CSR diffractometer, PD-3 at Dhruva reactor, Trombay, using neutrons at a wavelength of 1.48Å [6]. Scanning electron microscopy (SEM) image has been captured by Zeiss ultra-field emission SEM.

### Magnetic characterization

Temperature dependent *dc* magnetization (*M*) studies were carried out with the help of a commercial SQUID magnetometer (Quantum Design, USA) and *ac* $\chi$ study with different frequencies (v= 1.3, 13, 133, and 1333 Hz) with an *ac* field of 1 Oe was also carried out with the same magnetometer. Heat-capacity (C) studies were carried with a commercial Physical Properties Measurements System, PPMS (Quantum Design, USA)

### Dielectric and pyroelectric characterization

The PPMS system employed for heat-capacity studies was used to measure complex dielectric permittivity using an Agilent E4980A LCR meter with a home-made sample holder with several frequencies (1 kHz to 100 kHz) and with a bias voltage of 1 V; and the same sample holder was used for pyroelectric studies with Keithley 6517B electrometer by poling at 100 K with different electric fields.

In general, all measurements were performed for the zero-field-cooled condition (ZFC, from 300 K) of the specimen.

## Results and discussion

### Structural Studies

The structure of this compound was attributed by Laha et al. [2] to a rhombohedral space group $R\bar{3}m$ on the basis of XRD results (see Fig.1a). According to Ref. 2, in this structure, Li atoms occupy 3a site and Ni and Ru atoms occupy 3b site, forming layers of $LiO_6$ and $Ni(Ru)O_6$ octahedra running along c-axis. There is a further weak site-disorder with a small fraction of Li going to 3b site and Ni/Ru to 3a site. In Fig. 1b, we have shown the x-ray diffraction pattern for the title compound fitted by the LeBail method [7]. This XRD pattern seems to compare well with the one reported by Laha et al [2] and the lattice parameters deduced from the fits also are also comparable. Hence, it is reasonable to presume at this juncture that our synthesis has resulted in the same compound as obtained by Laha et al [2]. However, the neutron diffraction pattern recorded at 278 K using neutrons of wavelength 1.48Å shows that there are additional reflections (marked by asterisks) around 20.47º and 22.5º which are forbidden in space group $R\bar{3}m$. It has been observed that a number of Li compounds with layered rock-salt type structures crystallize in the monoclinic structure [1] and, therefore, the possibility of indexing the obtained neutron diffraction pattern using a monoclinic structure was explored. Using LeBail method, it is seen that the additional reflections can be properly indexed when the entire pattern is fitted in the monoclinic structure in the space group $P2/m$. Therefore, we have shown the fitting of XRD as well with this space group in figure 1b. The possible reasons for this ambiguity are: Firstly, x-ray atomic scattering form factors and neutron scattering lengths are quite different, resulting in widely different intensities for the same reflections, and secondly, the two peaks mentioned above are likely to arise from reflection planes dominated by lithium or oxygen atoms. It is well-known that x-ray scattering cross-section is very negligible for low atomic number elements like Li and O (compared to Ni and Ru), and the peaks dominated by scattering from these elements could be very feeble in XRD. Neutrons are ideal for identifying low Z elements due to the insensitivity of neutron scattering lengths to atomic number. The neutron scattering length values for Li, Ni, Ru and O are -0.19, 1.03 , 0.70, and 0.58 barns, respectively [8, 9]. Therefore, it can be reasonably assumed from the neutron diffraction data that the structure for $Li_3NiRuO_6$ could be described in a monoclinic structure in the space group $P2/m$, which is a sub-group of $R\bar{3}m$. We welcome future investigations aimed to confirm the inference made here on the nature of the sub-group. The homogeneity of the present compound was also investigated by SEM and the SEM image is shown in fig.1c. It is clear from the SEM image that all the grains are of similar size and we are not able to find any other impurity phase.

We make another observation from the temperature dependent neutron diffraction data. No additional reflections were observed as the temperature was lowered from 278K to 3 K. This rules out any long-range antiferromagnetic transition. Powder neutron diffraction profiles, measured at different temperatures, are plotted on an expanded scale to infer any variation in the intensities of the Bragg peaks as the temperature was gradually varied from 278 K to 3 K. In Fig. 2, the integrated intensity of the first strong Bragg peak (circled) is plotted as a function of temperature in the right inset. This peak was considered for the purpose of seeing the changes due to temperature variation, since this can be indexed in both rhombohedral as well as monoclinic space groups. With increasing temperature from 3 K, it is observed that there is a gradual change in the intensity of the first strong Bragg peak. The intensity peaks at around 20 K and then it falls as the temperature is raised. There is another anomaly around 100 K where the intensity rises again and then decreases on further increase of temperature. The temperatures at which the intensity profile shows anomalies are commensurate with the observations made from magnetization measurements, discussed below. It appears that short-range magnetic order inferred from the inverse $\chi$ plot (see below) is also sensed by the intensity increase below 200 K. We are not able to draw any conclusion on the magnetic structure from the present data.

**Magnetic Measurements**

*Magnetization:* The results of *dc* magnetic susceptibility as a function of *T* obtained in a field of *H* = 5 kOe are shown in Fig. 3a. There is a gradual increase of $\chi$ with decreasing *T* below 300 K down to 100 K, at which there is a sudden change in the slope of the curve; the upturn becomes sharper below about 80 K attributable to the onset of long-range magnetic ordering. There is a shoulder around 50 K, followed by a peak around 8 K, with a subsequent fall. These features suggest the existence of additional magnetic anomalies as the *T* is lowered below 80 K. For the sake of comparison with the results of Laha et al [2], the plot of inverse $\chi$ versus *T* is also shown in figure 3a. A linear region in the high temperature range (175 – 300 K) could be found. There is a deviation from the linearity as the temperature is lowered further, which is attributable to short-range magnetic correlations. Effective moment obtained from the Curie-Weiss fit from the linear region turns out to be about 4.71 $\mu_B$ per formula unit which is close to that expected (4.79 $\mu_B$) for divalent Ni (S= 1) and pentavalent Ru (S= 3/2). The value of the paramagnetic Curie-temperature is found to be about -374 K. These findings are in good agreement with those reported by Laha et al [2] by measurements with 1 kOe. We have carried out *dc* $\chi$ studies in a low field, that is, in 100 Oe, for ZFC and field-cooled (FC) conditions of specimen (see Fig. 3b). ZFC and FC curves show changes in slopes as in the curve obtained with 5 kOe in the magnetically ordered state, though qualitatively the shapes of the curves look different. What is clear is that the FC curve deviates from the ZFC curve below about 100 K, without any downfall. The fact remains that $\chi$(FC) continues to increase below 80 K, but the tendency to flatten appears below around 12 K only, and not at the onset of magnetic order. Irreversibility in ZFC-FC and the "delayed" flattening of the FC curve with multiple features as described above already signals complex magnetic nature. Though the irreversibility behaviour of ZFC-FC curves has been reported in many ferrimagnets and ferromagnets arising from magnetocrystalline anisotropy, there could be other possibilities - Griffiths phase or spin-glass freezing – as well. Therefore, the discussion below is focussed to address this issue. The feature around 12 K is rather sharp in these curves compared to that known for canonical spin-glasses, but such a sharpness is absent in the ac $\chi$ data (see below). This suggests that possible spin-glass freezing sensitively responds to a small application dc magnetic field.

We have measured hysteresis loops at low fields (Fig. 3c) and isothermal magnetization up to 140 kOe (Fig. 3d) at selected temperatures. *M(H)* at 1.8, 20 and 40 K (in the magnetically ordered state) continues to increase with *H* without any evidence for saturation at high fields. The plots at 1.8 and 20 K reveal weak hysteretic effect. There is a non-linearity of the curves at low fields below 40 K (see Fig. 3d) which is absent as one enters the paramagnetic state (say, at 150 K). Overall, these results reveal that the oxide is definitely not a ferromagnet in the magnetically ordered state.

In order to explore possible presence of Griffiths phase [10], we have obtained $\chi$(*T*) curves at several fields. A strong signature of Griffiths phase is that the sudden downturn in the plot of inverse $\chi$ at the onset of Griffiths phase would gradually get smeared with the application of *H*. We, however, note that the features look similar at low and high fields in the plot of inverse $\chi$, as shown in figure 4a. Therefore, the formation of Griffiths phase due to crystallographic disorder is ruled out in this material.

*Heat capacity measurements:* Figure 4b shows the behavior of $C(T)$ (in the absence of magnetic field) below 120 K. It is quite clear that heat-capacity decreases smoothly in the *T*-range of investigation without showing any peak or upturn attributed to long-range magnetic ordering of a ferromagnetic or antiferromagnetic-type. We have also plotted $C/T(T)$ in the inset of figure 4b to emphasize this point. Various explanations can be offered for the absence of any anomaly: (i) The entropy associated with the onset of magnetic ordering is small; (ii) the magnetic ordering temperature is high enough that the large lattice contribution masks any feature due to magnetic ordering; (iii) crystallographic disorder washes out the feature.

*Ac magnetization:* Figure 5a shows real ($\chi'$) and imaginary parts ($\chi''$) of *ac* $\chi$. With lowering of temperature, at the onset of magnetic ordering near 80 K, a weak upturn followed by a gradual increase is observed. There is no $\upsilon$-dependence of the feature around 80 K. However, $\chi''$ is featureless. Such a behaviour of $\chi'$ and $\chi''$ is not a signature [11] of spin-glass (for the transition around 80 K). Viewing together with non-saturation behavior of $M(H)$ at high fields, one can infer that the transition around 80 K is of a ferrimagnetic-type. However, with further lowering of temperature (in the absence of an external magnetic field), there is a peak in both these parts around 12 K for each frequency and the temperature at which this peak appears is also $\upsilon$-dependent (see also the inset of figure 5); the magnitude of the shift (about 1 K for a variation from 1.3 Hz to 1333 Hz) is typical [11] of many spin-glasses. In the presence of a *dc* magnetic field, say, of 5 kOe, the peaks are suppressed. These observations strongly favour the onset of a spin-glass regime around 12 K in this material. Thus, *ac* susceptibility data provide evidence for a re-entrant spin-glass behavior in this oxide.

We have analysed the *ac* $\chi$ results in terms of the conventional power law, $\tau/\tau_0 = (T_f/T_g - 1)^{-zv}$ [11]. Here, $\tau$ represents the observation time ($1/2\pi\upsilon$), $\tau_0$ is the microscopic relaxation time, $T_g$ is the spin-glass transition temperature, $T_f$ corresponds to freezing temperature for a given observation time and zv is the critical exponent. For the feature around 12 K, we obtained $T_g \approx 9.8$ K, $zv \approx 4.05$ and $\tau_0 \approx 5.7 \times 10^{-7}$ sec. The value of $\tau_0$ obtained is higher than that for the conventional spin glasses ($10^{-12}$s range), which implies cluster-glass behaviour below 12 K. We believe that there are clusters of ferrimagnets created by crystallographic disorder, which are randomly frozen.

*Isothermal remnant magnetization ($M_{IRM}$):* We measured $M_{IRM}$ at four selected temperatures, 1.8, 20, 40 and 150 K. After the specimen was zero-field-cooled to desired temperature, a field of 5 kOe was applied, and waited for some time before turning off the field. $M_{IRM}$ was then measured as a function of time ($t$). $M_{IRM}$ drops to small values within seconds of reducing the field to zero, but at 20 and 40K, the magnitude of $M_{IRM}(0)$ becomes much smaller when compared to that at 1.8K; in addition, it decays more slowly with $t$ at 1.8 K, as shown in figure 5b. At 150 K, the value of $M_{IRM}(0)$ is negligible. [The $M_{IRM}(0)$ values at t=0 for 1.8, 20, 40 and 150 K are 0.2688, 0.1634, 0.1472 and 0.00015 emu/g respectively]. These offer support to spin-glass dynamics well below 20 K only. The curve at 1.8 K could be fitted to a stretched exponential form of the type, $M_{IRM}(t) = M_{IRM}(0)[1+A\exp(-t/\tau)^{1-n})$, where A and n are constants and $\tau$ here is the relaxation time. It is found that the value of n and relaxation time $\tau$ are 0.42 and 53 minutes respectively. These values are in fact in agreement with that reported for cluster spin-glasses [11].

*Memory effect in dc magnetization:* In order to look for memory effect [12], we obtained $\chi(T)$ curves in different ways. In addition to ZFC curve in the presence of 100 Oe without a long wait at any temperature (which is a reference curve), we have obtained a ZFC curve after waiting at two temperatures 5 and 40 K for 3 hours each. The difference between these two curves is plotted (as $\Delta M$ versus $T$) in figure 5c. It is obvious that there is a clear sharp 'dip' at 5 K, which is below the $T_g$ (~12 K). However, there is no well-defined dip at 40 K. We repeated the measurements with a long waiting time ($t_{wait}$=6 hrs), and it was found that there is an increase in the intensity in $\Delta M$ at 5 K, while there is no change in intensity at 40 K. This conclusively establishes re-entrant spin-glass freezing in this material.

**Complex permittivity and pyroelectric behaviour**

Complex permittivity and pyroelectric current studies reveal interesting behaviour well below and above $T_g$ (~12 K). Dielectric constants ($\varepsilon'$) and the loss factor ($\tan\delta$) as a function of temperature are shown in figure 6a below 150 K, measured with a frequency of 100 kHz. We took the data with several frequencies (5, 10, 20, 30, 50, 70 and 100 kHz), but we did not see any frequency dependence. The values of $\tan\delta$ below 100 K are rather small and hence extrinsic contributions to dielectric permittivity can be ignored. Beyond 100 K, $\tan\delta$ increases and therefore extrinsic contributions tend to dominate. The observation we would like to stress is that both $\varepsilon'$ and $\tan\delta$ undergo a gradual increase with $T$ from 1.8 K, without any apparent peak or any other feature in zero magnetic field. However, when the complex permittivity was measured in the presence of a magnetic field ($H$= 50 kOe), a deviation of the curves around 60 - 80 K in $\varepsilon'$ and $\tan\delta$ with respect to zero-field curves is observed (see also the insets of figures 6a and 6b), suggestive of magnetodielectric coupling.

$I_{pyro}$ exhibits a distinct feature (measured with a poling voltage of 200 V corresponding to the electric field of 4.16 kV/cm) around 60 K (figure 6c), supporting the existence of an electric dipole anomaly. In addition, the plot shows another peak at about 40 K for a rate of warming of temperature ($dT/dt$) of 2K/min. The peak gets reversed, but asymmetrical, when poled by an electric field of opposite polarity. We have also obtained the behaviour of $I_{pyro}$ in the presence of $dc$ magnetic field of 50 kOe (see Fig. 6c) and two observations could be made: (i) The intensity of the peak in the plot is suppressed dramatically at 40 K and (ii) the hump, centred around 60 K, is resolved into a peak (as shown in the left inset of figure 6c). This is the same temperature where we see the deviation in the complex permittivity data in the presence of a magnetic field. When we reverse the electric field in the presence of 50 kOe magnetic field, the feature obtained near 60 K also gets reversed as shown in the right inset of figure 6c. In short, the dielectric data and the pyroelectric data reveal the presence of magneto-electric effect in this oxide. The observed pyroelectric peaks cannot arise from ferroelectricity, as there is no peak in the dielectric data in zero field. An alternate explanation to explain $I_{pyro}$ peaks can be proposed in terms of 'thermally stimulated depolarization current (TSDC)', the origin of which could lie in the crystallographic disorder as discussed in our earlier work [14-16].

In order to further explore magnetodielectric coupling, we have performed isothermal $\varepsilon'$ vs $H$ measurements at different temperature (1.8, 5, 14, 20, 30 and 40 K) up to 140 kOe. Figure 7 shows the $\Delta \varepsilon'$, where $\Delta \varepsilon' = [((\varepsilon'(H) - \varepsilon'(0)*100)/ \varepsilon'(0))]$ measured with a high frequency 100 kHz. It is observed that there is a sign reversal from negative to positive in MDE behaviour, when the sample is cooled from ferrimagnetic state to spin glass state. [The values at 40 K are less negative than that at 30 K deviating from the trend seen in comparison with the curves at other temperatures, and the difference is so negligible that we do not want to attach significance to this at this moment]. This further supports the existence of the coupling between the spin and the electric dipoles in this compound.

## Conclusions

We find re-entrant cluster spin-glass behaviour in a Li-based polycrystalline, $Li_3NiRuO_5$, not paid much attention for its properties in the literature. This oxide undergoes ferrimagnetic ordering below 80 K and enters spin glass regime around 12 K. Though there is no well-defined feature in the complex dielectric permittivity in the magnetically ordered state in the absence of an external magnetic field indicative of the absence of ferroelectricity, there is a distinct evidence for magnetodielectric coupling. Consistent with MDE coupling, the sign of MDE coupling also changes as the sample is cooled from ferrimagnetic state to spin-glass regime. There are pyroelectric anomalies in the vicinity of $30 - 70$ K, presumably arising from thermally stimulated depolarization current.

## References

1. S. Laha, E. Morán, R. Sáez-Puche, M. Á. Alario-Franco, A. J. Dos santos-Garcia, E. Gonzalo, A. Kuhn, F. García Alvarado, T. Sivakumar, S. Tamilarasan, S. Natarajan and J. Gopalakrishnan, *J. Solid State Chem.*, 2013, **203,** 160–165.
2. S. Laha, E. Morán, R. Sáez-Puche, M. Á. Alario-Franco, A. J. Dos santos-Garcia, E. Gonzalo, A. Kuhn, S. Natarajan, J. Gopalakrishnan and F. Garcia-Alvarado, *J. Mater. Chem. A*, 2013, **1**, 10686.
3. S. Laha, E. Morán, R. Sáez-Puche, M. Á. Alario-Franco, A. J. Dos santos-Garcia, E. Gonzalo, A. Kuhn, S. Natarajan, J. Gopalakrishnan and F. García-Alvarado *Phys. Chem. Chem. Phys*., 2015, **17,** 3749.
4. J. Lee, A. Urban, X. Li, D. Su, G. Hautier and G. Ceder, *Science*, 2014, **343**, 519–522.
5. B. C. Melot and J. M. Tarascon, *Acc. Chem. Res.*, 2013, **46**, 1226–1238.
6. V. Siruguri, P. D. Babu, M. Gupta, A. V. Pimpale, P. S. Goyal, Pramana - J Phys, 2008, 71, 1197 – 1202.
7. A. Le Bail, *Powder Diffraction*, 2005, **20**, 316-326.
8. J. Rodríguez-Carvajal, *Commission on Powder Diffraction (IUCr) Newsletter*, 2001, **26**, 12-19
9. J. Rodríguez-Carvajal, *Physica B*, 1993, **192**, 55-69.
10. R. B. Griffiths, *Phys. Rev. Lett.*, 1969, **23,** 17; A.H. Castro Neto, G. Castilla, and B. A. Jones, *Phys. Rev. Lett.*, 1998, **81,** 3531; W. Jiang, X.Z. Zhou, G. Williams, Y. Mukovskii, and K. Glazyrin, *Phys. Rev. Lett.*, 2007, **99,** 177203; E.V. Sampathkumaran, Niharika Mohapatra, Sudhindra Rayaprol, and Kartik K Iyer, *Phys. Rev. B*, 2007, **75**, 052412; Venkatesh Chandragiri, Kartik K Iyer and E.V. Sampathkumaran, *J. Phys. Condens. Matter*, 2016, **28,** 286002 and references therein.
11. J. A. Mydosh, *Spin glasses: An experimental introduction,* Taylor & Francis, 1993.
12. M. Vasilakaki, K.N. Trohidou, D. Peddis, D. Fiorani, R. Mathieu, M. Hudl, P. Nordblad, C. Binns, and S. Baker, *Phys. Rev. B,* 2013, **88,** 140402(R).
13. See for a review, W. Kleeman, *Solid State Phenom.,* 2012, **189,** 41-56.
14. Tao Zou, Zhiling Dun, Huibo Cao, Mengze Zhu, Daniel Coulter, Haidong Zhou, and Xianglin Ke, *Appl. Phys. Lett.,* 2014, **105,** 052906. Y. Kohara, Y. Yamasaki, Y. Onose, and Y. Tokura, *Phys. Rev. B,* 2010 **82,** 104419.
15. Sanjay Kumar Upadhyay, P.L. Paulose, Kartik K Iyer, and E.V. Sampathkumaran, Phys. Chem. Chem. Phys. **18,** 23348 (2016); Sanjay Kumar Upadhyay, Kartik K Iyer, S. Rayaprol, P.L. Paulose, and E.V. Sampathkumaran, Sci. Rep. **6,** 31883 (2016).
16. C. De, S. Ghara, and A. Sundaresan, Solid State Commun. **205,** 61 (2015).

## Figure captions

**Figure 1**
(**a**) Rhombohedral structure of $Li_3NiRuO_5$ compound.(b) LeBail fitting for powder x-ray diffraction (XRD) pattern (at 300 K) and neutron diffraction (ND) data (at 278 K) of $Li_3NiRuO_5$ assuming (top) monoclinic structure space group $P2/m$, and (bottom) rhombohedral structure space group $R\bar{3}m$. (c) SEM micrograph of $Li_3NiRuO_5$.

**Figure 2**:
Powder neutron diffraction data for $Li_3NiRuO_5$ measured at various temperatures is shown in an expanded form in the range $15-35^o$. Left inset highlights the first strong peak. In right inset, the normalized intensity of this peak is plotted as a function of temperature to highlight the variation of its intensity. The peak intensity has been normalized to the 278K data.

**Figure 3**:
For $Li_3NiRuO_5$, $dc$ magnetic susceptibility as a function of temperature measured in a magnetic field of 5 kOe, and 100 Oe are plotted in (a) and (b) respectively. In (a), inverse susceptibility is also plotted with the line through the data points showing the Curie-Weiss fitting. In (b), the curves obtained for ZFC and FC conditions are shown. (c) Hysteresis loops at 1.8 and 20 K with the inset showing the low-field region in an expanded form for 20 K. (d) Isothermal magnetization at 1.8, 20, 40 and 150 K.

**Figure 4**:
a) Inverse magnetic susceptibility (for ZFC condition) obtained with various magnetic fields for $Li_3NiRuO_5$. (b) Heat-capacity in zero magnetic field. Inset shows the $C/T$ plot.

**Figure 5**:
Real and imaginary parts of $ac$ susceptibility for $Li_3NiRuO_5$. The arrows show the direction in which the peaks shift with increasing frequency (1.3, 13, 133 and 1333 Hz). (b) Isothermal remnant magnetization at 1.8, 20 and 40 K, and (c) the difference in magnetization curves, $\Delta M$, obtained with and without waiting at 5 and 40 K for $Li_3NiRuO_5$. In figure (a), $ac$ data in the presence of a $dc$ magnetic field of 5 kOe are also shown. Inset in (a) shows the curves in an expanded temperature region around spin-glass transition.

**Figure 6**:
Temperature dependence of (a) dielectric constant ($\varepsilon'$), and (b) loss factor ($\tan\delta$) with and without external magnetic field of 50 kOe, measured with a frequency of 100 kHz for $Li_3NiRuO_5$. (c) Pyroelectric current, $I_{pyro}$, obtained by increasing the temperature at the rate of 2K/min after poling by 200 V for $Li_3NiRuO_5$. Inset of (a) and (b) show the curves below 100 K in an expanded form to show the change in curvature after the application of magnetic field. In (c), the curve obtained in a field of 50 kOe is also included. Left inset of (c) presents the zoomed curves near 60 K, while the right inset of (c) shows the zoomed curves around 60 K for positive and negative poling in a magnetic field of 50 kOe.

**Figure 7:**
Isothermal magneto-dielectric curves presented in the form of $\Delta\varepsilon'$ for selected temperatures for $Li_3NiRuO_5$.

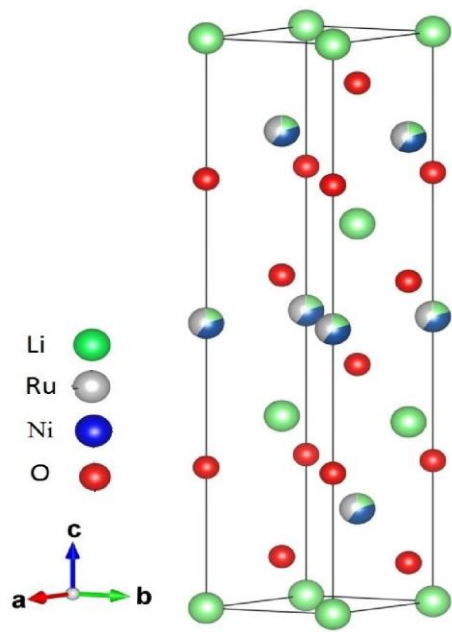

**Figure 1a**

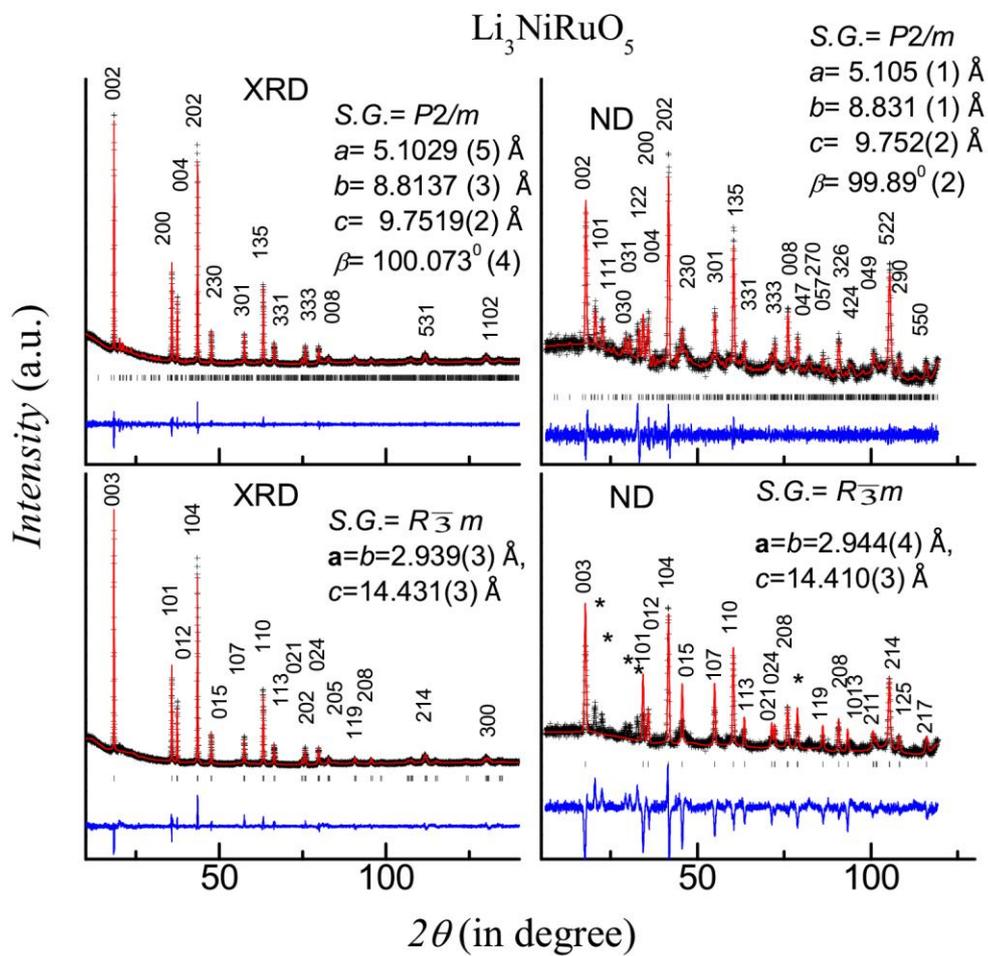

Figure 1b

Li₃NiRuO₅

XRD — S.G.= P2/m, a= 5.1029 (5) Å, b= 8.8137 (3) Å, c= 9.7519(2) Å, β= 100.073°(4)

ND — S.G.= P2/m, a= 5.105 (1) Å, b= 8.831 (1) Å, c= 9.752(2) Å, β= 99.89°(2)

XRD — S.G.= R$\bar{3}$m, a=b=2.939(3) Å, c=14.431(3) Å

ND — S.G.= R$\bar{3}$m, a=b=2.944(4) Å, c=14.410(3) Å

$2\theta$ (in degree)

Intensity (a.u.)

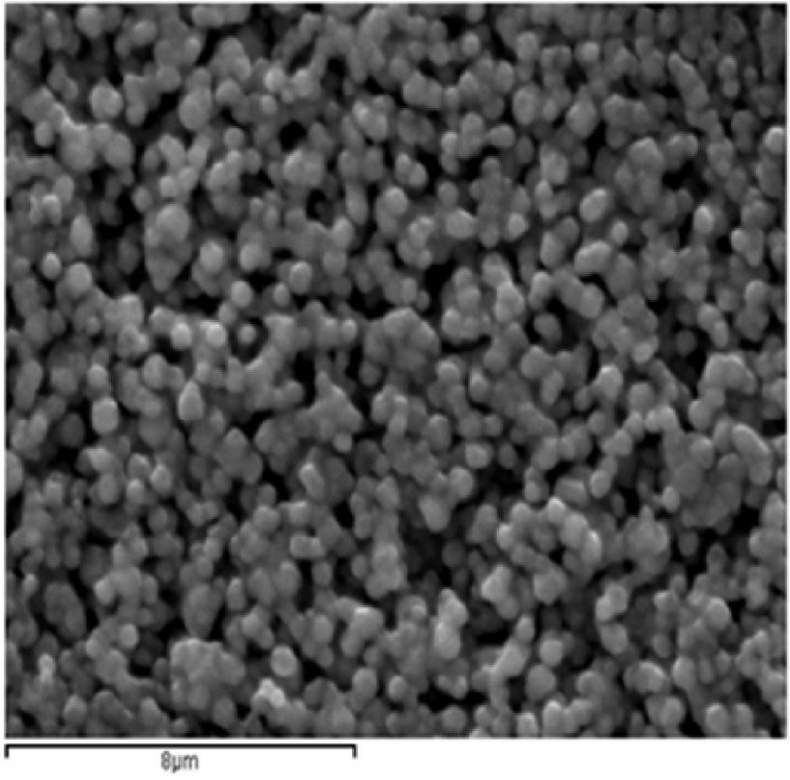

8μm

**Figure 1c**

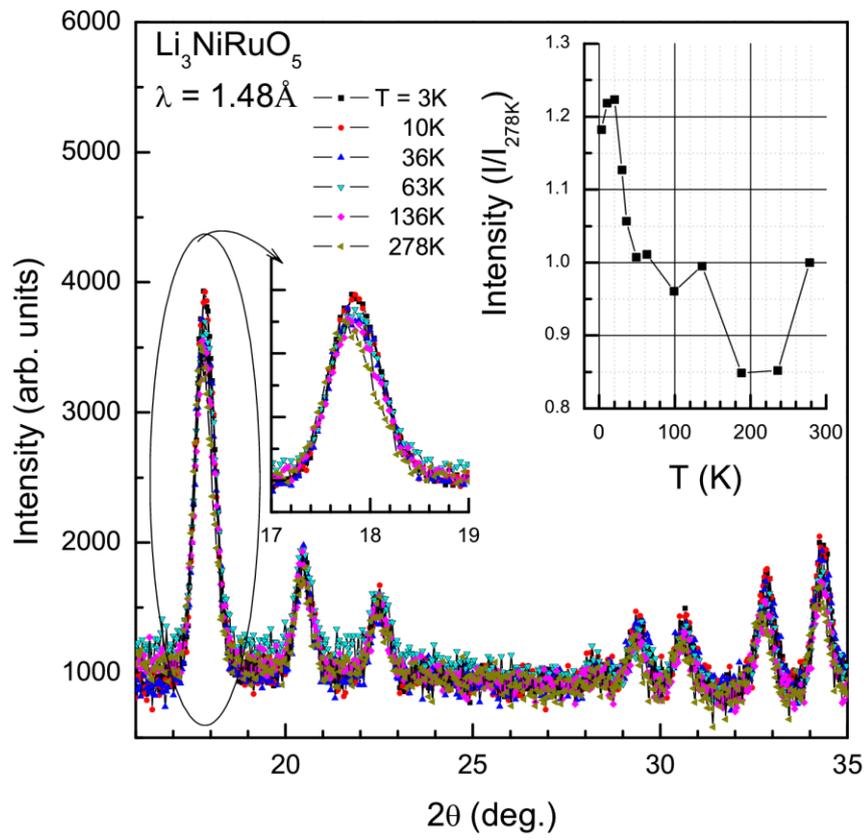

**Figure 2**

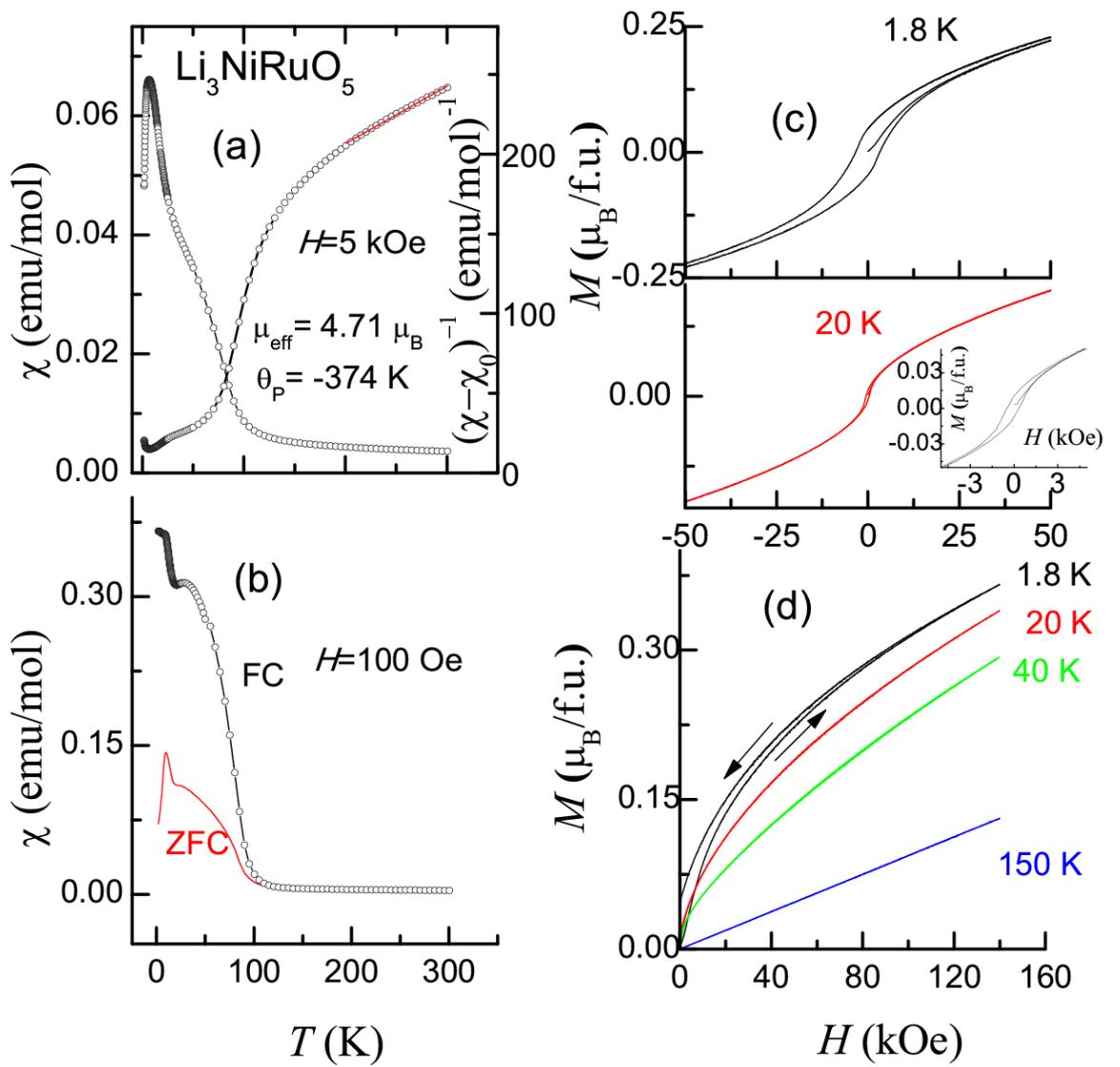

**Figure 3**

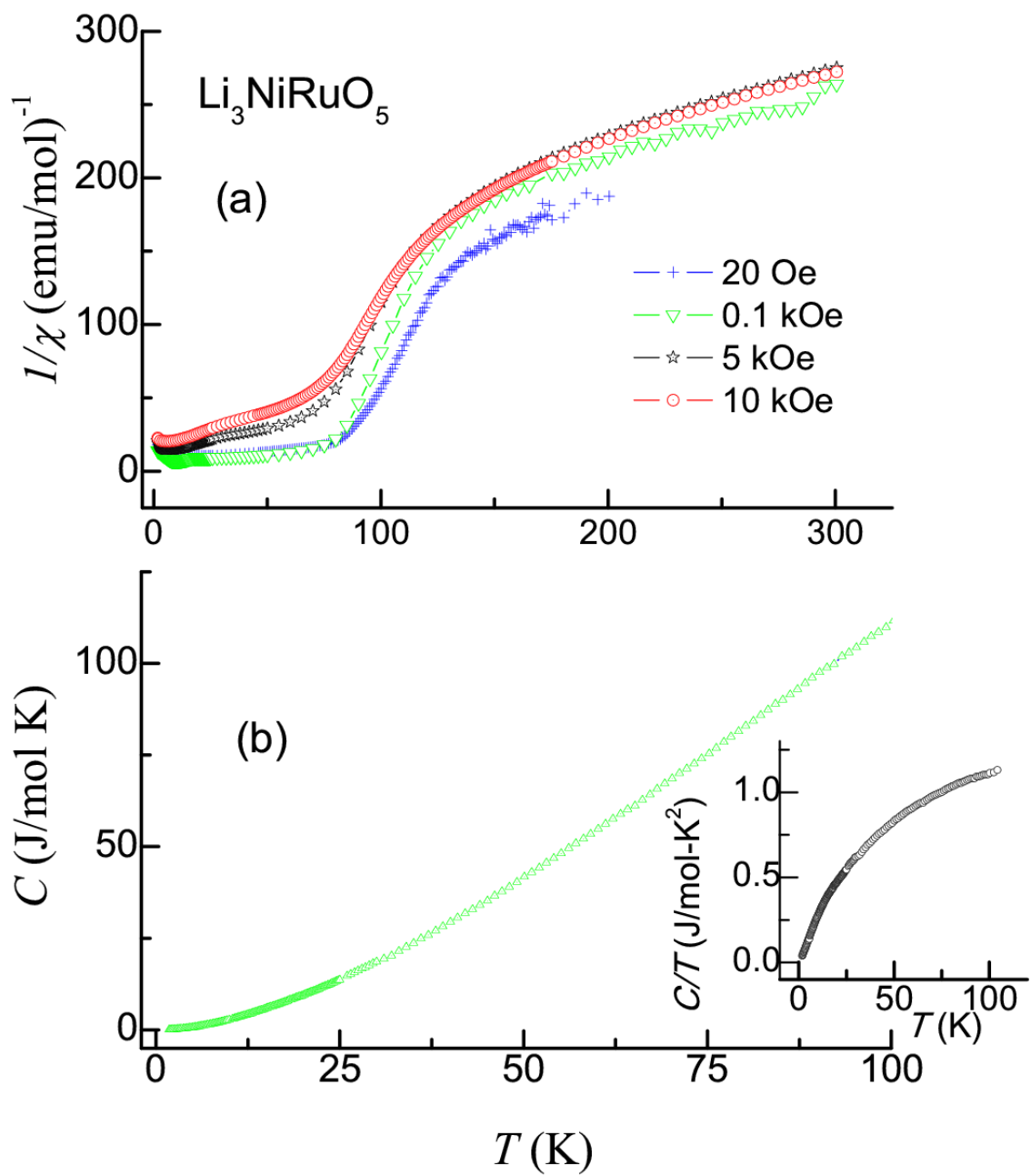

Figure 4

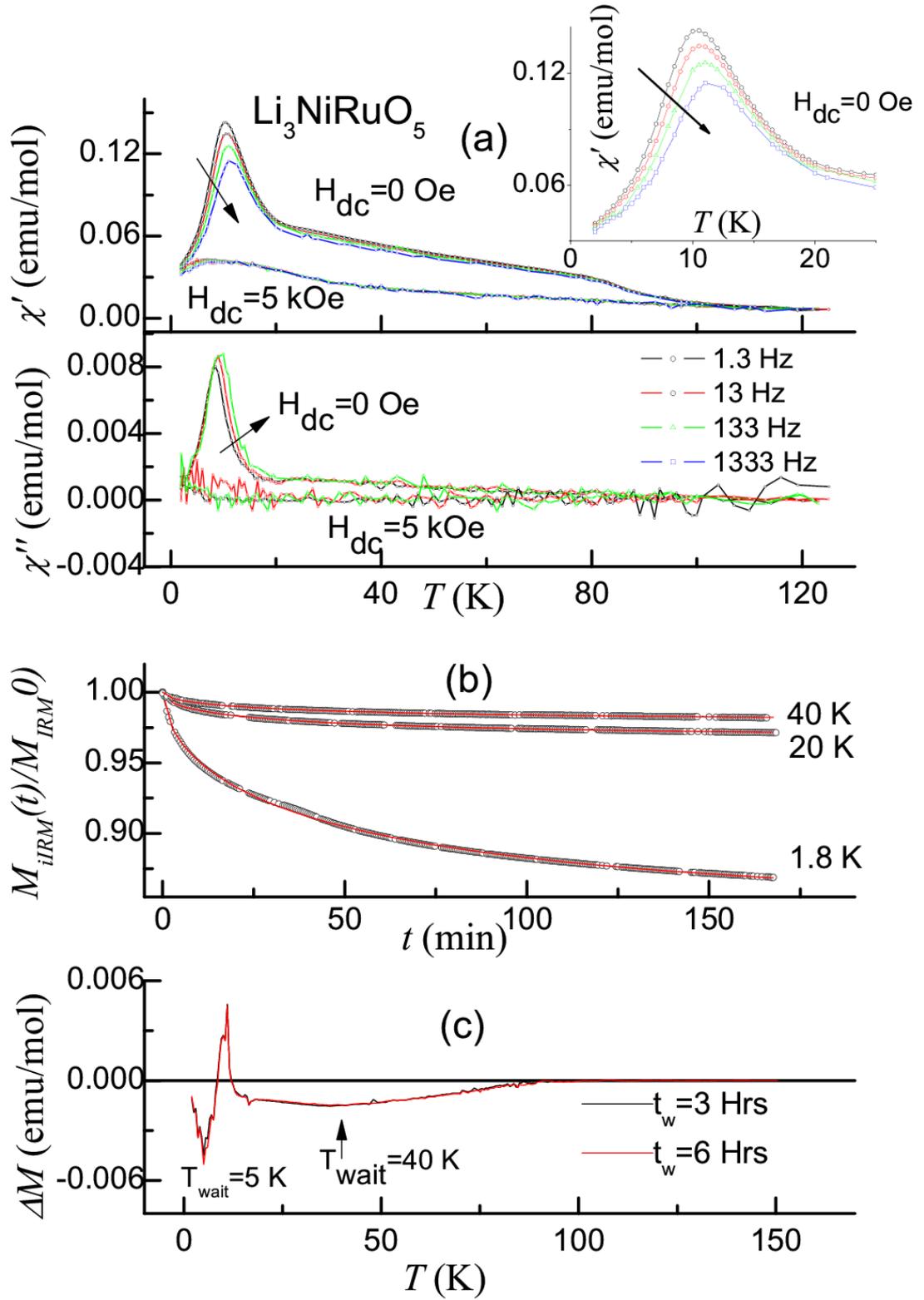

**Figure 5**

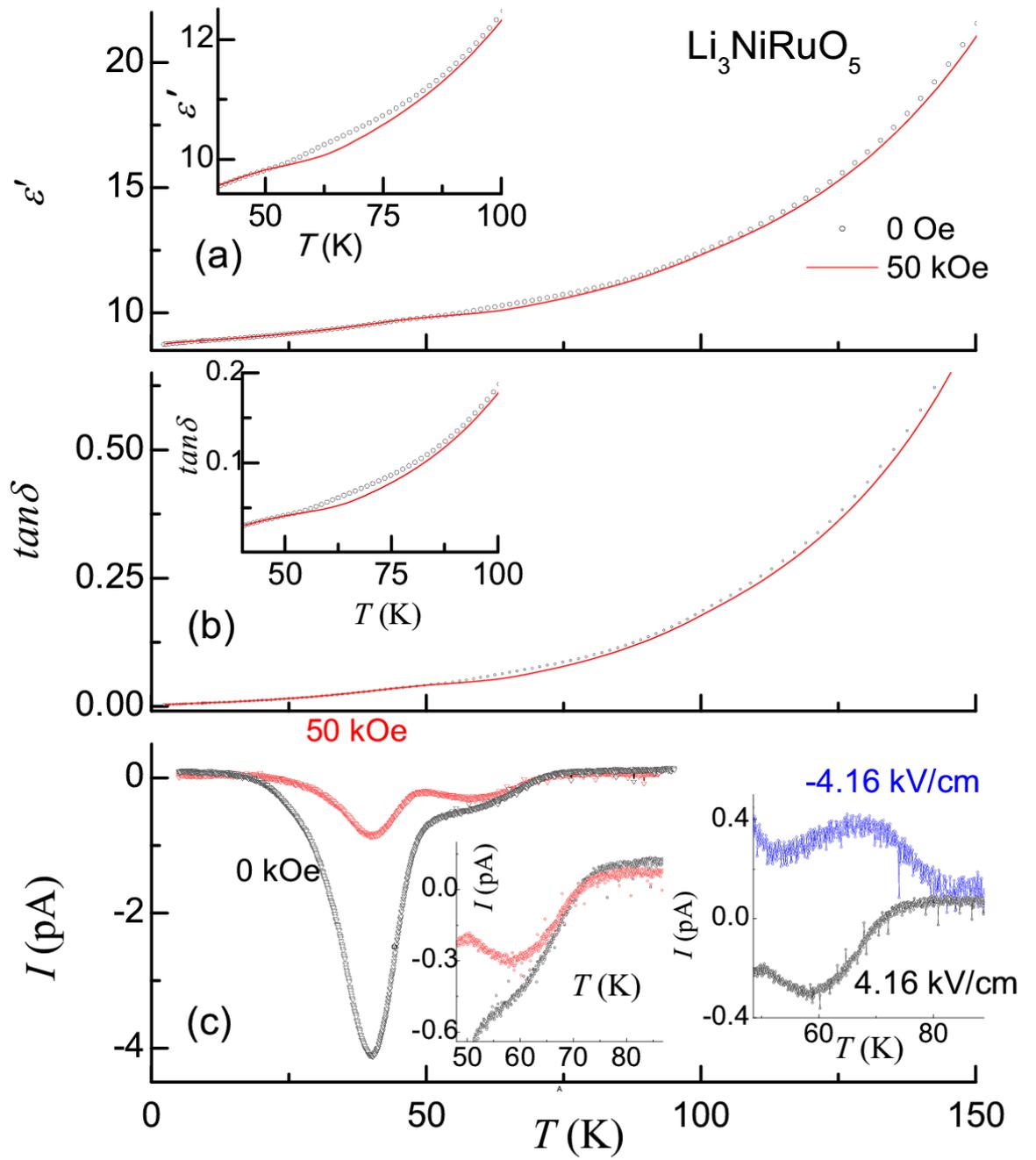

Figure 6

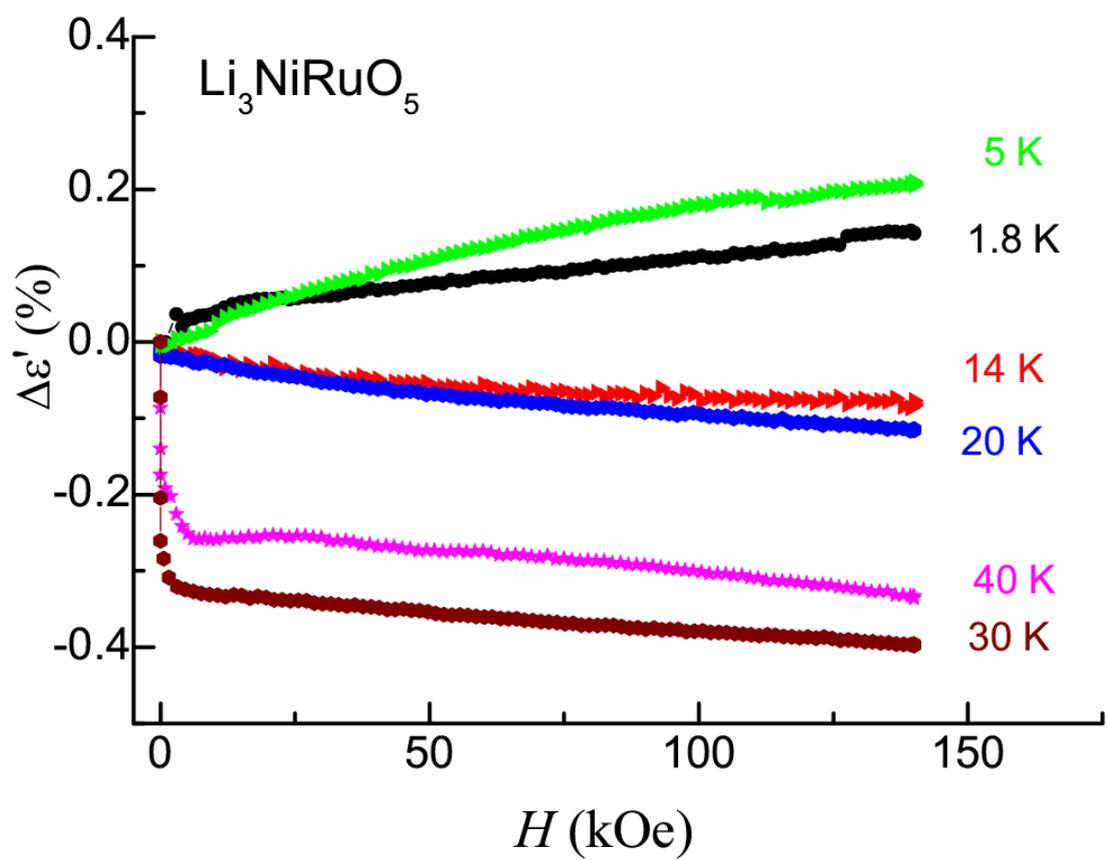

**Figure 7**